\documentclass[conference]{IEEEtran}
\IEEEoverridecommandlockouts

\usepackage{amsmath,amssymb,amsfonts}
\usepackage{caption}
\usepackage{subcaption}
\usepackage{algorithm}
\usepackage[justification=centering]{caption}
\usepackage{textcomp}
\usepackage{cite}
\usepackage{xcolor}
\usepackage{algorithm}
\usepackage{amsmath,amssymb}
\usepackage{epsfig}
\usepackage{setspace}
\usepackage{graphicx}

\usepackage{bm}
\usepackage[noend]{algpseudocode}
 
\usepackage{xcolor,soul,framed} 
\usepackage{algorithm}
\colorlet{shadecolor}{yellow}

\graphicspath{{../pdf/}{../jpeg/}}
\DeclareGraphicsExtensions{.pdf,.jpeg,.png}
\usepackage{url}
\usepackage{eqparbox}
\def\BibTeX{{\rm B\kern-.05em{\sc i\kern-.025em b}\kern-.08em
    T\kern-.1667em\lower.7ex\hbox{E}\kern-.125emX}}
\begin{document}

\title{A Fast Effective Greedy Approach for MU-MIMO Beam Selection in mm-Wave and THz Communications
\thanks{\authorrefmark{1}Indicates equal contribution to this work.}
}



\author{\authorblockN{Rafid Umayer~Murshed\authorrefmark{1}\authorrefmark{3}, Md Saheed~Ullah\authorrefmark{1}\authorrefmark{2}, and~Mohammad Saquib\authorrefmark{3} }

\authorblockA{Email: rafidumayer.murshed@utdallas.edu, saheed@udel.edu, saquib@utdallas.edu}
 \authorblockA{\authorrefmark{3}Department of Electrical and Computer Engineering, The University of Texas at Dallas, Texas, USA}
  \authorblockA{\authorrefmark{2}Department of Electrical and Computer Engineering, University of Delaware, Delaware, USA}}

\maketitle

\begin{abstract}
This paper addresses the beam-selection challenges in Multi-User Multiple Input Multiple Output (MU-MIMO) beamforming for mm-wave and THz channels, focusing on the pivotal aspect of spectral efficiency (SE) and computational efficiency. We introduce a novel approach, the Greedy Interference-Optimized Singular Vector Beam-selection (G-IOSVB) algorithm, which offers a strategic balance between high SE and low computational complexity. Our study embarks on a comparative analysis of G-IOSVB against the traditional IOSVB and the exhaustive Singular-Vector Beamspace Search (SVBS) algorithms. The findings reveal that while SVBS achieves the highest SE, it incurs significant computational costs, approximately 162 seconds per channel realization. In contrast, G-IOSVB aligns closely with IOSVB in SE performance yet is markedly more computationally efficient. Heatmaps vividly demonstrate this efficiency, highlighting G-IOSVB's reduced computation time without sacrificing SE. We also delve into the mathematical intricacies of G-IOSVB, demonstrating its theoretical and practical superiority through rigorous expressions and detailed algorithmic analysis. The numerical results illustrate that G-IOSVB stands out as an efficient, practical solution for MU-MIMO systems, making it a promising candidate for high-speed, high-efficiency wireless communication networks.
\end{abstract}


\section{Introduction}
\IEEEPARstart{T}{he} advancement of wireless communication technologies, particularly massive multiple-input multiple-output (MIMO) systems, plays a pivotal role in addressing the increasing demand for high-throughput and reliable connections \cite{lu2014overview,Raf_URLLC}. These systems, characterized by their dense antenna arrays, are crucial in enhancing spectral efficiency (SE) and mitigating interference, especially in 5G and 6G networks \cite{MIMO_6G_ref}. In the context of Multi-User MIMO (MU-MIMO) systems, which cater to elevated data rates and network capacity, managing inter-user interference emerges as a formidable challenge \cite{intro_MU_MIMO}. This has led to significant research in beamforming algorithms, particularly focusing on beam-selection strategies to optimize user-specific signal paths \cite{Beam_selection_intro}. Despite numerous advancements in beamforming, including low-complexity precoding, spatial division-multiplexing \cite{adhikary2013joint}, and coordinated beamforming \cite{bjornson2010cooperative}, computational complexity remains a persistent issue. Recent developments in beam-selection techniques, such as adaptive and predictive beam-selection methods, have shown promise in enhancing computational efficiency \cite{Beam_Selection_Algo1,Beam_Selection_Algo2}.

The transition to terahertz (THz) frequencies in beamforming, offering vast bandwidth and data rates, poses new challenges, especially in beam-selection due to THz-specific channel properties \cite{UM_MIMO_THz_algos1,ullah2022spectral}. These challenges necessitate innovative approaches to beam-selection in THz frequencies, as seen in recent explorations \cite{THz_Beam_selection}. Furthermore, the hybrid precoding in millimeter-wave (mmWave) systems, balancing hardware complexity and SE, faces scalability challenges in large MIMO systems \cite{yu2017hybrid, nguyen2016hybrid}. Here, efficient beam-selection strategies are crucial for reducing computational overhead while maintaining performance \cite{MM_wave_Beam_selection}.

Addressing these complexities, our work introduces the Greedy Interference-Optimized Singular Vector Beam-selection (G-IOSVB) algorithm. Focused on MU-MIMO systems, this approach is designed to minimize inter-user interference, enhance SINR and SE, and simplify beam-selection processes, marking a significant advancement in wireless communication optimization.

We concentrate on a downlink MU-MIMO beamforming system with a base station (BS) and multiple users, emphasizing beam-selection. Our proposed low-complexity algorithm efficiently manages inter-user interference to achieve high SE. The contributions of this paper include:

\begin{itemize}
\item A novel analytical framework redefining cumulative inter-user interference in MU-MIMO systems through a mathematically rigorous approach, including lemmas, a corollary, and a theorem.
\item The introduction of a low-complexity greedy algorithm for MU-MIMO beamforming, with an emphasis on beam-selection, and an exploration of its theoretical computational advantages.
\item Numerical analysis validating the performance of the proposed greedy algorithm, with a focus on computational efficiency and SE. The analysis includes parameter optimization across diverse MU-MIMO scenarios, demonstrating its efficiency compared to traditional beamforming methods.
\end{itemize}

The rest of the paper is organized as follows. MU-MIMO system, including channel modeling, is described, and the problem is formulated in Section II. The proposed beam-selection approach is presented in Section III. We depict and analyze the numerical results in Section IV. Section V concludes the paper.

This paper adopts the following notation conventions: Scalars are denoted by lowercase letters, while bold upper and lower case letters represent matrices and vectors, respectively. The norms for scalars, vectors, and matrices are symbolized as $|.|$ for scalar norms, $||.||$ for the $L_2$ norms of vectors, and $||.||_F$ for Frobenius norms of matrices. The transpose and conjugate transpose of a matrix or vector \textbf{x} is indicated by $\mathbf{x^\top}$ and \textbf{x\textsuperscript{$H$}}, respectively. Furthermore, $\mathbb{C}$ represents the set of complex numbers, and $\mathbb{Z}^+$ denotes the set of positive integers.

\section{System Model and Problem Formulation}
In this section, we describe the MU-MIMO beamforming system and channel model. We then formulate the basic multi-user beamforming problem.

\subsection{System Model}

We consider a downlink MU-MIMO system with $U$ users, each equipped with $N_r$ receive antennas. The base station (BS), featuring $N_t$ transmit antennas, serves $N_s$ data streams to each user. We focus on simplified beam-selection strategies and do not employ hybrid beamforming architectures. However, any recently developed efficient beamforming approaches \cite{self_HBF} can be employed for practical, real-life systems after beam-selection.

The transmitted signal at the BS, $\textbf{x}$, is a function of the digital beamforming matrix $\textbf{F}_D$ and the data symbol vector $\textbf{s}$ of length $UN_s$. The digital beamforming matrix $\textbf{F}_D$ is a $N_t \times UN_s$ matrix, with each user's digital beamforming component represented as $\textbf{F}_{k}^D$, a sub-matrix of dimensions $N_t \times N_s$. The channel matrix for the $k^{th}$ user is denoted as $\textbf{H}_k$, having dimensions $N_r \times N_t$. 

\subsection{Channel Model}

In our study, we employ two distinct channel models to evaluate the performance of our algorithms: the Saleh-Valenzuela (SV) model for mmWave bands and the NYUSIM model for THz bands. The SV model, a cluster-based statistical approach, effectively simulates mmWave band characteristics, essential for modern wireless systems\cite{ullah2023beyond}. On the other hand, NYUSIM is tailored for the THz band, especially in indoor scenarios, incorporating comprehensive statistical data for realistic emulation of THz communication complexities\cite{ju2023142}. Utilizing both models ensures our algorithms' adaptability and effectiveness across a range of wireless communication environments, aligning with the diverse frequency spectrum used in contemporary wireless technologies.

\subsection{Problem Formulation}

The challenge in MU-MIMO systems lies in optimizing SE while managing interference efficiently. While some iteration-based algorithms address interference \cite{alkhateeb2015limited,yuan2018hybrid}, their computational efficiency is often limited. Therefore, our objective is to maximize SE with effective interference management in a time-efficient manner.

We define the optimal beam-selection matrices as $\textbf{F}^\text{io}$ and $\textbf{W}^\text{io}$, which contain the beam directions minimizing interference between users. Here, $\textbf{F}^\text{io}$ is a $N_t \times UN_s$ matrix, representing transmit beam directions, and $\textbf{W}^\text{io}$ is a $N_r \times UN_s$ matrix, representing receive beam directions. Each $\textbf{F}_k^\text{io}$, a sub-matrix of $\textbf{F}^\text{io}$, is a $N_t \times N_s$ matrix for the $k$-th user, and similarly, $\textbf{W}_k^\text{io}$ is a $N_r \times N_s$ matrix for the $k$-th user.

The data rate for this system model can be expressed as:
\begin{equation}
R_D=\sum_{k=1}^{U} \log_2(1+\frac{||{(\textbf{W}_k^\text{io})}^H \textbf{H}_k \textbf{F}_k^\text{io} ||^2}{\Delta^D_k + ||{(\textbf{W}_k^\text{io} )}^H \textbf{n}_k||^2}),
\end{equation}
where $\mathbf{H}_k\in \mathbb{C}^{N_r \times N_t}$ denotes the channel matrix and $\Delta^D_k = \sum_{\substack{i=1 \\ i \neq k}}^{U} \left\| \left(\mathbf{W}_k^\text{io}\right)^H \mathbf{H}_k \mathbf{F}_i^\text{io} \right\|^2$, the interference for the $k$-th user.

The primary problem is to find the matrices $\textbf{F}_k^\text{io}$ and $\textbf{W}_k^\text{io}$ that minimize interference and maximize the sum rate. The optimization task involves selecting $\textbf{F}^\text{io}$ and the corresponding $\textbf{W}^\text{io}$ from all different possible combinations.

\section{Methodology}
We commence this section by providing a concise overview of the benchmarking algorithms introduced in \cite{ullah2023beyond}. The proposed G-IOSVB algorithm is subsequently detailed. Precise mathematical expressions are formulated to determine the number of iterations necessary for the G-IOSVB algorithm, and the theoretical computational advantage over the original IOSVB algorithm is deduced.

\subsection{Benchmark Algorithms}
Prior to exploring the intricate nuances of the G-IOSVB algorithm under consideration, we begin with a brief synopsis of the algorithms that were developed in \cite{ullah2023beyond}.
\subsubsection{SVBS Algorithm}
The Singular-Vector Beamspace Search (SVBS) Algorithm, developed in \cite{ullah2023beyond} for MU-MIMO, serves as a tool to benchmark the upper limits of achievable SE. However, its exhaustive nature renders it impractical for real-world applications due to the significant computational burden it imposes.

\subsubsection{IOSVB Algorithm}
The Interference Optimized Singular Vector Beamforming (IOSVB) algorithm \cite{ullah2023beyond} is a sophisticated approach designed for MU-MIMO systems, focusing on optimizing fully digital beamforming matrices $\textbf{F}^\text{io}$ and $\textbf{W}^\text{io}$. This method diverges from exhaustive search techniques by instead identifying near-optimal beamforming matrices from candidate matrices $\textbf{F}^\text{sel} \in \mathbb{C}^{N_t \times UR_{\mathrm{sel}}}$ and $\textbf{W}^\text{sel}\in \mathbb{C}^{N_r \times UR_{\mathrm{sel}}}$ reducing computational complexity.

The candidate beamforming matrices are constructed by selecting the first $R_\text{sel} (\geq N_s)$ columns from the SVD of user channel matrices, forming $\textbf{F}^\text{sel}$ and $\textbf{W}^\text{sel}$. The channel matrix of the $k$-th user can be decomposed as $\textbf{U}_k\mathbf{\Sigma}_k \textbf{V}_k^H$. The first $R_\text{sel}$ singular values of each user are then concatenated to obtain $\mathbf{\Sigma}$. The correlation matrix $\textbf{C}^\text{sel}\in \mathbb{C}^{UR_{\mathrm{sel}} \times UR_{\mathrm{sel}}}$ captures the interplay between beam directions across users:

\begin{equation}\label{csel}
    \textbf{C}^\text{sel} = (\textbf{F}^\text{sel} \;\mathbf{\Sigma}^H)^H \times \textbf{F}^\text{sel}.
\end{equation}

For each specific beam combination selected with indexes $\textbf{ind}\in (\mathbb{Z}^+)^{UN_s}$,  a specific correlation matrix $\textbf{C}_\text{corr}^i\in \mathbb{C}^{UN_s \times UN_s}$ is defined, determining the interaction of selected beam directions:
 \begin{equation}
     \textbf{C}_\text{corr}^i = (\textbf{F}^\text{sel}[\textbf{ind}]\mathbf{\Sigma}^H[\textbf{ind}])^H \times \textbf{F}^\text{sel}[\textbf{ind}].
 \end{equation}

The IOSVB algorithm is rooted in the concept that system interference, particularly between users, is closely related to the correlation of their beam directions. The interference for user $k$ is expressed as:
\begin{equation}
\Delta^D_k = \sum_{i=1, i\neq k}^{U}\left\| \left(\mathbf{W}_k^\text{io}\right)^H \mathbf{H}_k \mathbf{F}_i^\text{io} \right\|^2.
\end{equation}

Corollary 1, Lemma 1, and Lemma 2 (see appendix for proof) collectively form the mathematical foundation of the IOSVB algorithm. They show that interference can be represented in terms of singular vectors and values:
\begin{equation}
    \left\| \left(\mathbf{W}_k^\text{io}\right)^H \mathbf{H}_k \mathbf{F}_i^\text{io} \right\|^2 = \left\| \bm{\Sigma}_k[\mathbf{ind}_k] \mathbf{V}_k^H \mathbf{V}_i[\mathbf{ind}_i] \right\|^2.
\end{equation}

The total interference $I_t$ across all users is then expressed as the sum of squared off-diagonal elements of $\textbf{C}_\text{corr}^i$:
\begin{equation}
\resizebox{.9\hsize}{!}{$
I_t = \sum_{k=1}^U \sum_{\substack{i=1 \\ i \neq k}}^U \left\| \left(\mathbf{W}_k^\text{io}\right)^H \mathbf{H}_k \mathbf{F}_i^\text{io} \right\|^2 = \left\| \mathbf{C}_{\text{corr}}^i - \text{diag}(\mathbf{C}_{\text{corr}}^i) \right\|^2,
$}
\end{equation}

This theorem (see appendix for proof), the core of the IOSVB algorithm, implies that minimizing the off-diagonal elements of $\textbf{C}_\text{corr}^i$ reduces total system interference, guiding the selection of optimal beamforming matrices for enhanced system performance. A more comprehensive description of the implementation of IOSVB can be found in \cite{ullah2023beyond}.

\subsection{Proposed Greedy-IOSVB Algorithm}

In our proposed novel G-IOSVB algorithm, we refine the selection of the $ \textbf{C}_\text{corr}^i$ matrix through a greedy iterative process over $N_s$ steps. Unlike the exhaustive approach that simultaneously evaluates all singular vector orientation combinations, our method strategically selects an optimal subset of U columns from the $\mathbf{C}^\text{sel}$ matrix in each iteration. This process, excluding previously chosen columns, iteratively reduces the search space. Initially, it considers $\binom{R\text{sel}}{{1}}^U$ column combinations, which progressively diminish to $\binom{R_\text{sel}-m}{{1}}^U$ in the $m$-th iteration. This sequential refinement substantially trims the computational complexity, making it a more efficient alternative to the exhaustive $\binom{R_\text{sel}}{{N_s}}^U$ search paradigm.

The objective function remains:

\begin{equation}
f= {{||\mathbf{C}_\text{corr}-\mathrm{diag(\mathbf{C}_\text{corr})}||}_F}.
\end{equation}

We maintain an index set $\textbf{ind}$, which is updated iteratively to record the indices of the selected columns. During the m-th greedy phase, the correlation matrix $\textbf{C}_\text{corr}^i$ for the i-th combination is calculated by masking the non-selected columns in $\mathbf{C}^\text{sel}$ with a boolean matrix $\mathbf{M}_\text{sel}^i$, where the entries corresponding to the selected columns are one, and all other entries are zero. The correlation matrix is updated as follows:

\begin{equation}
\textbf{C}_\text{corr}^i = \textbf{C}^\text{sel} \circ \textbf{M}_\text{sel}^i; \forall i \in {1,2,....,{\binom{R_\text{sel}-m+1}{1}}^U},
\end{equation}

where $m$ denotes the current greedy iteration, and the Hadamard product is used to perform element-wise multiplication.

The selection of the optimal U columns at each iteration is subject to the following constraints:

\begin{align}
\sigma_\text{sel} &> \gamma \sigma_\text{max}, \\
\textbf{C}_\text{corr} &\in \left\{\textbf{C}_\text{corr}^1, \textbf{C}_\text{corr}^2, \ldots, \textbf{C}_\text{corr}^{\binom{R_\text{sel}-m+1}{1}^U}\right\},
\end{align}

where, $\gamma$ is a predefined channel gain threshold, $0<\gamma<1$, $\sigma_\text{max}$ indicates the maximum possible received power, and $\sigma_\text{sel}$ is the sum of the selected channel gains \cite{ullah2023beyond}.

After completing $N_s$ iterations of greedy selection, the final set of selected column indices is contained within $\mathbf{ind}$. The optimal fully digital precoder and combiner matrices $\mathbf{F}^\text{io}$ and $\mathbf{W}^\text{io}$ are then determined by extracting the columns from $\mathbf{C}^\text{sel}$ corresponding to the indices in $\mathbf{ind}$, yielding the minimum value of the objective function across all iterations. The analytical implementation of G-IOSVB is summarized in Algorithm \ref{alg_g-iosvb}.

\subsection{Number of Iterations}

In the development of the G-IOSVB algorithm, a critical aspect to consider is the total number of iterative searches required for optimizing the selection of column vectors from the \( \mathbf{C}^\text{sel} \) matrix. This process fundamentally deviates from the original IOSVB, which searches across all \( \binom{R_\text{sel}}{N_s}^U \) combinations, instead employing a more computationally efficient iterative approach.

Initially, the total number of iterative searches, $N_\text{iter}^G$, required for the G-IOSVB could be represented as:

\begin{equation}
N_\text{iter}^G = \sum_{m=1}^{N_s} \binom{R_\text{sel} - m + 1}{1}^U.
\end{equation}

This expression can be simplified and understood more deeply through a series of mathematical manipulations. Recognizing that \(\binom{R_\text{sel} - m + 1}{1}\) simplifies to \(R_\text{sel} - m + 1\), we reformulate the expression as:

\begin{equation}
N_\text{iter}^G = \sum_{m=1}^{N_s} (R_\text{sel} - m + 1)^U.
\end{equation}

By employing the binomial theorem, this sum of powers of binomials is further expanded into:

\begin{equation}
\resizebox{.89\hsize}{!}{$
N_\text{iter}^G = \sum_{m=1}^{N_s} \sum_{i=0}^{U} \binom{U}{i} R_\text{sel}^{(U-i)} (-1)^i (m - 1)^i.
$}
\end{equation}

This expression can be further simplified by applying Faulhaber's formula, which provides a closed-form expression for the sum of powers of natural numbers:

\begin{equation}
\resizebox{.89\hsize}{!}{$
N_\text{iter}^G = \sum_{i=0}^{U} \binom{U}{i} R_\text{sel}^{(U-i)} (-1)^i \left[ \frac{1}{i + 1} \sum_{j=0}^{i} \binom{i + 1}{j} B_j (N_s)^{i + 1 - j} \right].$
}
\end{equation}

\begin{equation}
\resizebox{.85\hsize}{!}{$
\begin{aligned}
N_\text{iter}^G &= \sum_{i=0}^{U} \binom{U}{i} R_\text{sel}^{(U-i)} (-1)^i \times \\
           &\quad \left[ \frac{1}{i + 1} \left( B_0 N_s^{i+1} + \binom{i+1}{1} B_1 N_s^{i} + \dots + \binom{i+1}{i} B_i \right) \right].
\end{aligned}
$}
\end{equation}

where \( B_j \) are the Bernoulli numbers. This sum in $N_{\text{iter}}^G$ is shaped by the interplay between its terms: the dominant $B_0 N_s^{i+1}$ component, scaling directly with $N_s$, and the fine-tuning higher-order Bernoulli numbers, $B_j$ for $j > 0$, which gain significance as $N_s$ increases. The variable $U$ expands the summation range, adding complexity. This nuanced calculation demonstrates the algorithm's ability to manage complexity, especially in large-scale scenarios with substantial $R_\text{sel}$ and $U$. Striking a balance between computational feasibility and efficiency, G-IOSVB transitions from exhaustive to iterative search, significantly reducing computational demands without compromising performance, a crucial aspect for real-world applications.

\subsection{Theoretical Computational Gain}

This section explores the theoretical computational gain of the Greedy-IOSVB (G-IOSVB) algorithm compared to the traditional IOSVB approach. The gain is quantified by comparing the total number of iterations required by both methods. 

\begin{equation}
\resizebox{.88\hsize}{!}{$
\frac{N_{\text{iter}}}{N_{\text{iter}}^G} = \frac{\left( \binom{R_{\text{sel}}}{N_s} \right)^U}{\sum_{i=0}^{U} \binom{U}{i} R_{sel}^{(U-i)} (-1)^i \left[ \frac{1}{i + 1} \sum_{j=0}^{i} \binom{i + 1}{j} B_j (N_s)^{i + 1 - j} \right]}
$}
.\end{equation}

To elucidate the computational gain, we employ Stirling's approximation for factorials and condense the binomial coefficients. For ease of computation and practicality, we consider only the first two Bernoulli numbers ($B_0 = 1$ and $B_1 = - 1/2$). The ratio of iterations between G-IOSVB and IOSVB is approximated as:

\begin{equation}
\resizebox{.89\hsize}{!}{$
\frac{N_{\text{iter}}}{ N_{\text{iter}}^G} \approx \frac{\left( \frac{\sqrt{2\pi R_{\text{sel}}} \left( \frac{R_{\text{sel}}}{e} \right)^{R_{\text{sel}}}}{\sqrt{2\pi N_s} \left( \frac{N_s}{e} \right)^{N_s} \sqrt{2\pi (R_{\text{sel}} - N_s)} \left( \frac{R_{\text{sel}} - N_s}{e} \right)^{R_{\text{sel}} - N_s}} \right)^U}{\sum_{i=0}^{U} \frac{U^i}{i!} R_{\text{sel}}^{(U-i)} (-1)^i \left[ \frac{1}{i + 1} \left(N_s^{i+1} - \frac{1}{2} \binom{i + 1}{1} N_s^i \right) \right]}.
$}
\end{equation}

The G-IOSVB algorithm's computational advantage over the original IOSVB becomes more pronounced with increases in \( R_{\text{sel}} \), data streams (\( N_s \)), and user count (\( U \)). A higher \( R_{\text{sel}} \) amplifies the numerator's growth, highlighting the exhaustive search's heavier computational load. Rising \( N_s \) values add complexity but emphasize G-IOSVB's efficiency. Most notably, an increase in \( U \) exponentially heightens the computational disparity between the two methods. These elements demonstrate G-IOSVB's growing effectiveness and substantial computational savings in handling larger-scale problems.

\begin{algorithm}
\caption{Greedy-IOSVB (G-IOSVB)}
\label{alg_g-iosvb}
\begin{algorithmic}[1]
\State Input:
$\{\bm{H}_k\}_{k=1}^U$, $N_s$, $R_{sel}$, $\gamma$
\State Initialize $\textbf{F}^{io}$ and $\textbf{W}^{io}$ as empty matrices
\State Set initial $f(0)=10^9$
\State Calculate $\sigma_{\text{max}} =\sum_{i=1}^{U}\sum_{j=1}^{N_s}\sigma_i^j$
\For {$m = 1$ to $N_s$}
\State Update $R_{\text{sel}}^{(m)} = R_{\text{sel}} - m + 1$
\State Run IOSVB algorithm with $N_s' = 1$, $R_{\text{sel}} = R_{\text{sel}}^{(m)}$
\State Concatenate the output beams to $\textbf{F}^{io}$ and $\textbf{W}^{io}$
\EndFor
\State Output: Optimized beamforming matrices $\textbf{F}^{io}$, $\textbf{W}^{io}$
\end{algorithmic}
\end{algorithm}

\begin{figure}
  \begin{center}
  \includegraphics[height=5.5cm,trim={2cm .3cm 2cm 1.6cm},clip,width=\columnwidth]{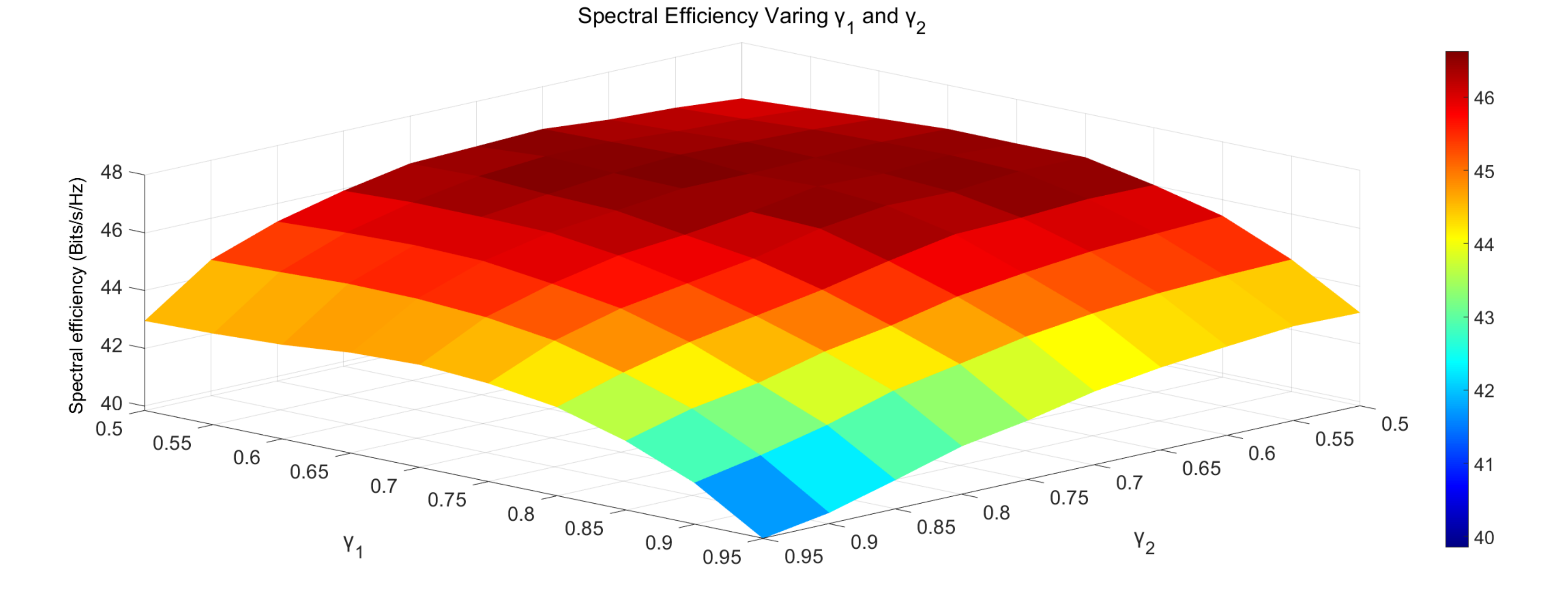}
  \caption{{Heatmap displaying SE variation across different values of channel gain thresholds $\gamma_1$ and $\gamma_2$.}}
  \label{fig_gamma_heatmap}
  \end{center}
\end{figure}

\section{Results}
This section outlines the simulation parameter configuration utilized to assess the performance of the proposed algorithm. Additionally, it demonstrates the procedure for determining the optimal algorithmic parameter configuration that optimizes the SE of the MU-MIMO system. Additionally, we analyze the proposed method's SE compared to established beamforming algorithms.

\subsection{Simulation Setup and Parameters}
MATLAB is employed to simulate each of the beamforming algorithms. The computational hardware for generating simulation results comprises a Core i7 processor, 32GB of RAM, and a 4GB GPU. Table \ref{overall_res} presents a comprehensive overview of the considered simulation parameters.

\begin{table}[ht]
\caption{Simulation Parameters.}
\centering
{
\begin{tabular}{cc}
\hline
Parameters          &             Values
\\ 
\hline
No. of Data Streams, $N_s$ & $2-6$
\\
No. of Propagation Paths, $L_k$ & 50
\\
Transmit Antennas, $N_t$ & 144
\\
Receive Antenna Per User, $N_r$ & $16$
\\
No. of Users, $U$ & $5$
\\
Channel Gain Ratio, $\gamma$ & 0.6
\\
No. of Selected Candidate Columns, $R_{sel}$ & 3-6
\\
\hline
\end{tabular}}
\label{overall_res}
\end{table}

\subsection{Selection of Algorithmic Parameters}
The $\gamma$ must be selected with care to optimize the proposed algorithm. Here, we delineate how these parameters can be determined via numerical analysis. In our study focusing on optimizing the G-IOSVB algorithm for $R_{sel} = 4$ and $N_s = 2$, we conducted simulations to determine the ideal settings for the parameters $\gamma_1$ and $\gamma_2$. As illustrated in Fig. \ref{fig_gamma_heatmap}, our findings indicate that setting $\gamma_1$ and $\gamma_2$ to approximately 0.6 yields the highest SE. According to this notable result, a uniform choice of $\gamma$ parameters across iterations can maximize SE. This simplified parameter selection of the G-IOSVB makes it easier to apply efficiently across different configurations of $R_{sel}$ and $N_s$, enhancing the algorithm's practicality and resilience.

\begin{figure}
     \centering
     \begin{subfigure}[b]{0.5\textwidth}
      \centering
         \includegraphics[height=5.5cm,trim={3cm .3cm 3.1cm 1cm},clip,width=3in]{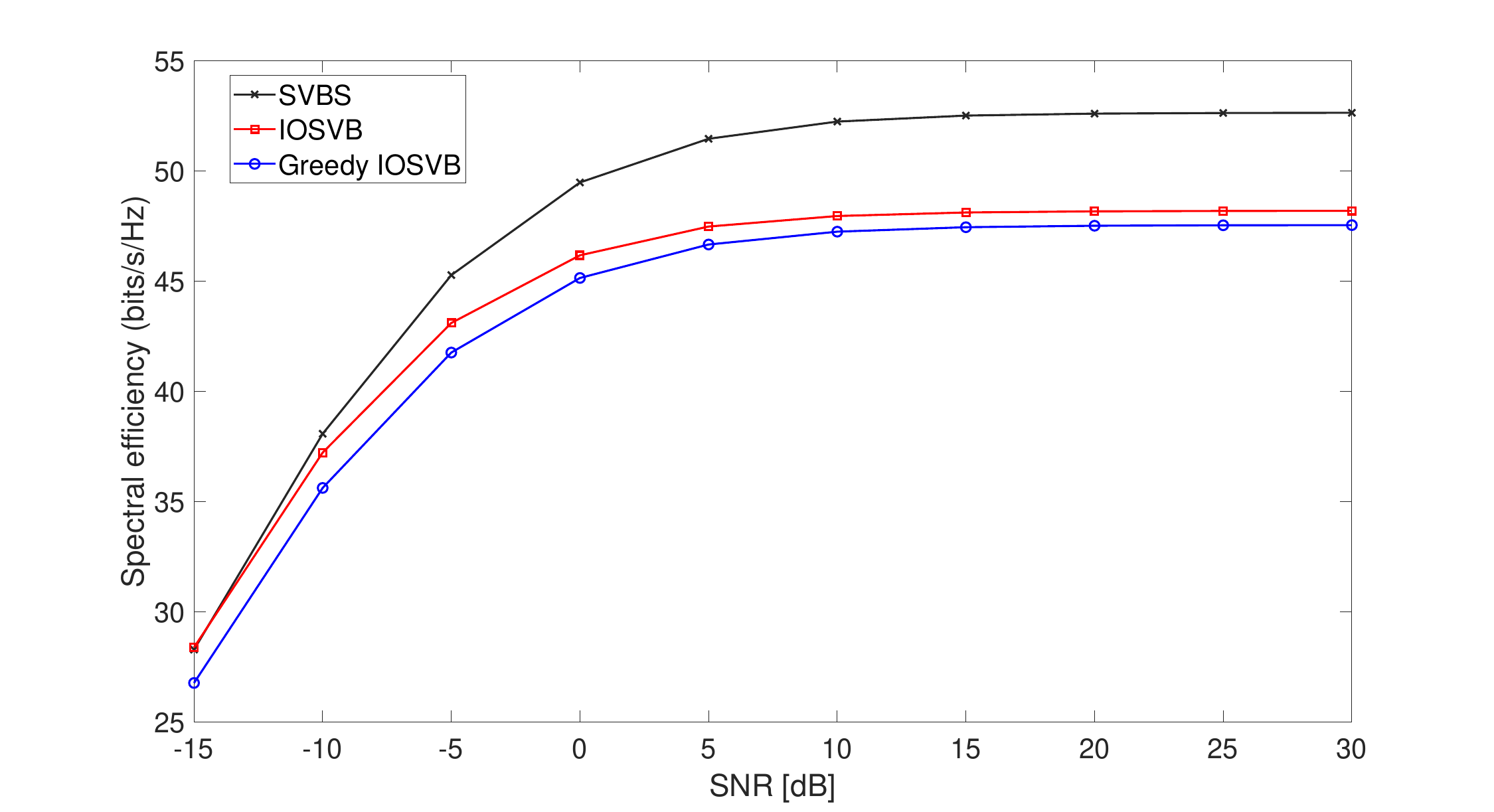}
         \caption{For mm-wave channels}
         \label{fig_SE_mmwave}
     \end{subfigure}
     \hfill
     \begin{subfigure}[b]{0.5\textwidth}
         \centering
         \includegraphics[height=5.5cm,trim={3cm .3cm 3cm 1cm},clip,width=3in]{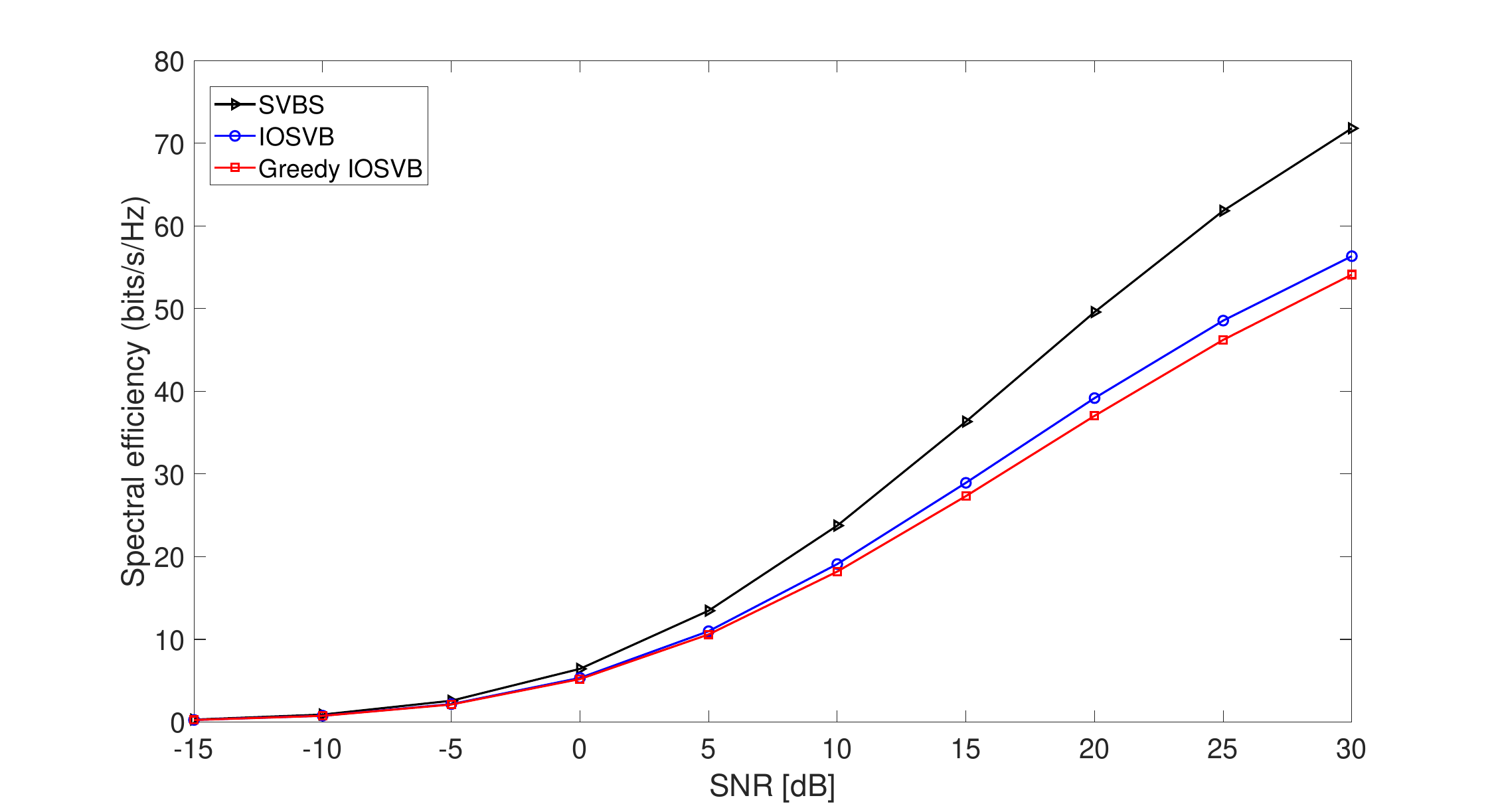}
         \caption{For THz channels}
         \label{fig_SE_THz}
     \end{subfigure}
        \caption{SE Performance: G-IOSVB vs. Traditional Algorithms in MU-MIMO Systems.}
        \label{fig_all_se_comp}
\end{figure}

\begin{table}[ht]
\centering
\caption{Performance comparison of IOSVB and G-IOSVB algorithms (at 25dB SNR).}
\label{table-comparison}
\scalebox{1}[1.15]{
\resizebox{\columnwidth}{!}{%
\begin{tabular}{|c|c|c|c|c|c|c|c|}
\hline
$\mathbf{N_s}$ & $\mathbf{R_{sel}}$ & \multicolumn{3}{c|}{\textbf{IOSVB}} & \multicolumn{3}{c|}{\textbf{G-IOSVB}} \\ \hline
            &               & $\mathbf{N_{iter}}$ & \textbf{Time (ms)} & \textbf{SE} & $\mathbf{N^G_{iter}}$ & \textbf{Time (ms)} & \textbf{SE} \\ \hline
2           & 4             & 7776           & 71.21              & 47.98                   & 1267            & 27.67              & 46.53                   \\ \hline
2           & 5             & $10^5$         & 401.51             & 49.06                   & 4149            & 51.21              & 47.47                   \\ \hline
3           & 5             & $10^5$         & 891.22             & 60.47                   & 4392            & 53.12              & 58.34                   \\ \hline
3           & 6             & $>10^6$        & 12893              & 61.39                   & 11925           & 90.59              & 59.06                   \\ \hline
\end{tabular}
}}
\end{table}

\subsection{Performance Analysis}

Our comprehensive analysis evaluates the SE of IOSVB, G-IOSVB, and SVBS algorithms. Figures \ref{fig_SE_mmwave} and \ref{fig_SE_THz} present the simulation results from 1000 channel realizations for mmWave and THz channels, respectively. SVBS achieves the highest SE but is computationally intensive, requiring approximately 162 seconds per realization. In contrast, G-IOSVB provides a similar level of SE with significantly reduced computational demands (requiring a few milliseconds), demonstrating its applicability and efficiency in both mmWave and THz environments.

\begin{figure}
  \begin{center}
  \includegraphics[trim={1.0cm 1.0cm 1.0cm 1.0cm},clip,width=\columnwidth]{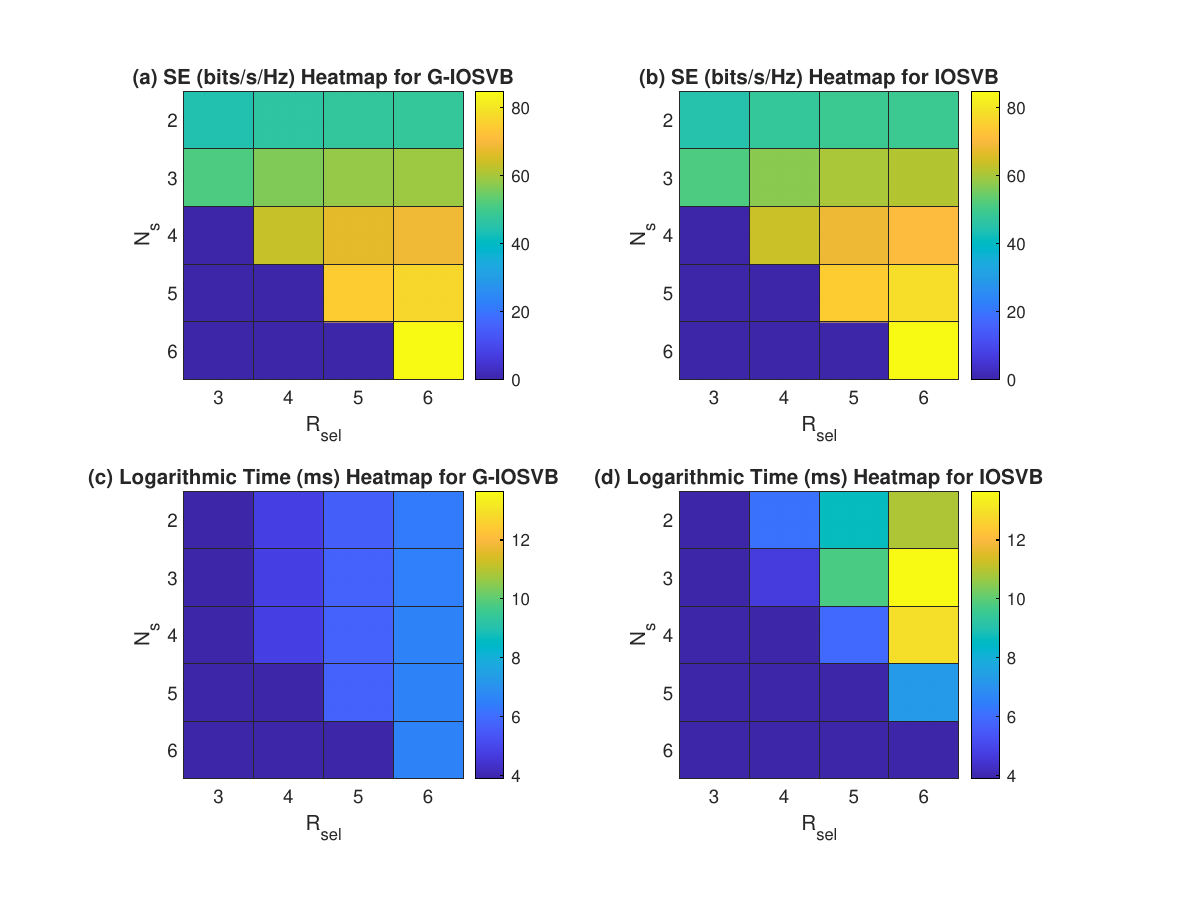}
  \caption{{Comparison of SE and computational time of G-IOSVB and IOSVB across different $R_{sel}$ and $N_s$ values.}}
  \label{Main_heatmap}
  \end{center}
\end{figure}

To further compare the G-IOSVB with traditional IOSVB, extensive simulations were conducted, and each algorithm executed over 1000 independent channel realizations by varying the values of $R_{sel}$ and $N_s$. The results, illustrated in Fig. \ref{Main_heatmap} through heatmaps, reveal G-IOSVB's capability to match IOSVB in SE across diverse settings of $R_{sel}$ and $N_s$. This alignment underscores G-IOSVB's effectiveness in maintaining high SE in MU-MIMO systems. However, G-IOSVB's superiority lies in its computational efficiency, which requires considerably less time than IOSVB to achieve comparable SE. Additionally, a summarized comparison of IOSVB and G-IOSVB algorithms is presented in Table \ref{table-comparison}, emphasizing the computational efficiency of G-IOSVB. Across various scenarios, G-IOSVB consistently demonstrates a lower computational time while maintaining comparable SE with IOSVB, further validating its advantage in practical applications. This empirical evidence reinforces G-IOSVB's role as a scalable and efficient approach for contemporary MU-MIMO systems.


\section{Conclusion}
This study presents the G-IOSVB algorithm, a significant advancement in MU-MIMO beam-selection focused on SE and computational efficiency. Through rigorous mathematical analysis and extensive simulations, G-IOSVB demonstrates a capability to maintain high SE, akin to traditional IOSVB and the exhaustive SVBS method, but with markedly reduced computational demands. The algorithm's efficiency, particularly in mmWave and THz frequencies, is evident in its comparative analysis with existing approaches. Thus, G-IOSVB is a practical, efficient solution for contemporary and future high-speed wireless communication systems, marking a notable contribution to the field. 

\appendix

\textit{Corollary 1:} If $\mathbf{U}_k$ is an orthonormal matrix, then the product of $\mathbf{U}_k$ and the submatrix $\mathbf{U}_k^H[\mathbf{ind}_k]$ results in a sparse matrix, which can be identified as a non-square identity matrix, i.e.,
\begin{equation}
    \mathbf{U}_k^H[\mathbf{ind}_k] \mathbf{U}_k = \mathbf{I}_k[\mathbf{ind}_k],
\end{equation}
where $\mathbf{I}_k$ is a non-square identity matrix.

\textit{Proof:} Given that $\mathbf{U}_k$ is an orthonormal matrix of size $N_r \times N_r$, the product $\mathbf{U}_k^H \mathbf{U}_k$ yields an $N_r \times N_r$ identity matrix. Since $\mathbf{U}_k^H[\mathbf{ind}_k]$ is of size $N_s \times N_r$, the product with $\mathbf{U}_k$ results in a non-square identity matrix $\mathbf{I}_k[\mathbf{ind}_k]$ of size $N_s \times N_r$, where non-matching index multiplications yield zero.

\textit{Lemma 1:} For an interfering user $i$ and a desired user $k$, the interference can be represented as:
\begin{equation}
    \left\| \left(\mathbf{W}_k^{io}\right)^H \mathbf{H}_k \mathbf{F}_i^{io} \right\|^2 = \left\| \bm{\Sigma}_k[\mathbf{ind}_k] \mathbf{V}_k^H \mathbf{V}_i[\mathbf{ind}_i] \right\|^2.
\end{equation}

\textit{Proof:} Decomposing $\mathbf{H}_k$ via singular value decomposition, we have:
\begin{equation}
\left\| \left(\mathbf{W}_k^{io}\right)^H \mathbf{H}_k \mathbf{F}_i^{io} \right\|^2 = \left\| \left(\mathbf{W}_k^{io}\right)^H \mathbf{U}_k \bm{\Sigma}_k \mathbf{V}_k^H \mathbf{F}_i^{io} \right\|^2.
\end{equation}
Applying \textit{Corollary 1}, we obtain:
\begin{equation}
\left\| \left(\mathbf{W}_k^{io}\right)^H \mathbf{H}_k \mathbf{F}_i^{io} \right\|^2 = \left\| \mathbf{I}_k[\mathbf{ind}_k] \bm{\Sigma}_k \mathbf{V}_k^H \mathbf{V}_i[\mathbf{ind}_i] \right\|^2.
\end{equation}

\textit{Lemma 2:} Given two matrices $\mathbf{A}$ and $\mathbf{B}$, with $\mathbf{A} = \begin{bmatrix} x_1 & \cdots & x_k & \cdots & x_U \end{bmatrix}^T$ and $\mathbf{B} = \begin{bmatrix} y_1 & \cdots & y_i & \cdots & y_U \end{bmatrix}$, the product excluding the diagonal can be expressed as:
\begin{equation}
\mathbf{A}^T \mathbf{B} - \text{diag}(\mathbf{A}^T \mathbf{B}) = \mathbf{A}^T \mathbf{B}_{i \neq k}.
\end{equation}

\textit{Proof:} The product $\mathbf{A} \mathbf{B}$ yields a matrix with diagonal elements $x_i y_i$. Excluding these diagonal elements, we obtain $\mathbf{A}^T \mathbf{B}_{i \neq k}$.

\textit{Theorem:} The total interference $I_t$ over all users, defined as $\sum_{k=1}^U \Delta^D_k = I_t$, can be expressed as:
\begin{equation}
\resizebox{.88\hsize}{!}{$
I_t = \sum_{k=1}^U \sum_{\substack{i=1 \\ i \neq k}}^U \left\| \left(\mathbf{W}_k^{io}\right)^H \mathbf{H}_k \mathbf{F}_i^{io} \right\|^2 = \left\| \mathbf{C}_{\text{corr}}^i - \text{diag}(\mathbf{C}_{\text{corr}}^i) \right\|^2,
$}
\end{equation}
where $\mathbf{C}_{\text{corr}}^i = \left(\mathbf{F}^{sel}[\mathbf{ind}] \bm{\Sigma}[\mathbf{ind}]\right)^H \mathbf{F}^{sel}[\mathbf{ind}]$ is the correlation matrix of the system.

\textit{Proof:} By applying \textit{Lemma 1} and \textit{Lemma 2}, and considering the structure of the correlation matrix $\mathbf{C}_{\text{corr}}^i$, the total interference $I_t$ can be simplified to the non-diagonal elements of $\mathbf{C}_{\text{corr}}^i$, thus proving the theorem.


\bibliographystyle{IEEEtran}
\bibliography{bibliographyy}

\begin{thebibliography}{10}
\providecommand{\url}[1]{#1}
\csname url@samestyle\endcsname
\providecommand{\newblock}{\relax}
\providecommand{\bibinfo}[2]{#2}
\providecommand{\BIBentrySTDinterwordspacing}{\spaceskip=0pt\relax}
\providecommand{\BIBentryALTinterwordstretchfactor}{4}
\providecommand{\BIBentryALTinterwordspacing}{\spaceskip=\fontdimen2\font plus
\BIBentryALTinterwordstretchfactor\fontdimen3\font minus \fontdimen4\font\relax}
\providecommand{\BIBforeignlanguage}[2]{{%
\expandafter\ifx\csname l@#1\endcsname\relax
\typeout{** WARNING: IEEEtran.bst: No hyphenation pattern has been}%
\typeout{** loaded for the language `#1'. Using the pattern for}%
\typeout{** the default language instead.}%
\else
\language=\csname l@#1\endcsname
\fi
#2}}
\providecommand{\BIBdecl}{\relax}
\BIBdecl

\bibitem{lu2014overview}
L.~Lu, G.~Y. Li, A.~L. Swindlehurst, A.~Ashikhmin, and R.~Zhang, ``An overview of massive mimo: Benefits and challenges,'' \emph{IEEE journal of selected topics in signal processing}, vol.~8, no.~5, pp. 742--758, 2014.

\bibitem{Raf_URLLC}
R.~U. Murshed, A.~Horaira~Hridhon, and M.~F. Hossain, ``Deep learning based power allocation in 6g urllc for jointly optimizing latency and reliability,'' in \emph{2021 5th International Conference on Electrical Information and Communication Technology (EICT)}, 2021, pp. 1--6.

\bibitem{MIMO_6G_ref}
F.~A. Pereira~de Figueiredo, ``An overview of massive mimo for 5g and 6g,'' \emph{IEEE Latin America Transactions}, vol.~20, no.~6, pp. 931--940, 2022.

\bibitem{intro_MU_MIMO}
\BIBentryALTinterwordspacing
D.~Gesbert, S.~Hanly, H.~Huang, S.~S. Shitz, O.~Simeone, and W.~Yu, ``Multi-cell mimo cooperative networks: A new look at interference,'' \emph{IEEE J.Sel. A. Commun.}, vol.~28, no.~9, p. 1380–1408, dec 2010. [Online]. Available: \url{https://doi.org/10.1109/JSAC.2010.101202}
\BIBentrySTDinterwordspacing

\bibitem{Beam_selection_intro}
Q.~Xue, X.~Fang, M.~Xiao, S.~Mumtaz, and J.~Rodriguez, ``Beam management for millimeter-wave beamspace mu-mimo systems,'' \emph{IEEE Transactions on Communications}, vol.~67, no.~1, pp. 205--217, 2019.

\bibitem{adhikary2013joint}
A.~Adhikary, J.~Nam, J.-Y. Ahn, and G.~Caire, ``Joint spatial division and multiplexing—the large-scale array regime,'' \emph{IEEE transactions on information theory}, vol.~59, no.~10, pp. 6441--6463, 2013.

\bibitem{bjornson2010cooperative}
E.~Bj{\"o}rnson, R.~Zakhour, D.~Gesbert, and B.~Ottersten, ``Cooperative multicell precoding: Rate region characterization and distributed strategies with instantaneous and statistical csi,'' \emph{IEEE Transactions on Signal Processing}, vol.~58, no.~8, pp. 4298--4310, 2010.

\bibitem{Beam_Selection_Algo1}
Q.~Zhang, X.~Li, B.-Y. Wu, L.~Cheng, and Y.~Gao, ``On the complexity reduction of beam selection algorithms for beamspace mimo systems,'' \emph{IEEE Wireless Communications Letters}, vol.~10, no.~7, pp. 1439--1443, 2021.

\bibitem{Beam_Selection_Algo2}
Y.~Wang, Q.~Li, J.~Jiao, S.~Wu, and Q.~Zhang, ``Arm: Adaptive random-selected multi-beamforming estimation scheme for satellite-based internet of things,'' \emph{IEEE Access}, vol.~7, pp. 63\,264--63\,276, 2019.

\bibitem{UM_MIMO_THz_algos1}
B.~Ning, Z.~Tian, W.~Mei, Z.~Chen, C.~Han, S.~Li, J.~Yuan, and R.~Zhang, ``Beamforming technologies for ultra-massive mimo in terahertz communications,'' \emph{IEEE Open Journal of the Communications Society}, vol.~4, pp. 614--658, 2023.

\bibitem{ullah2022spectral}
M.~S. Ullah, S.~C. Sarker, Z.~B. Ashraf, and M.~F. Uddin, ``Spectral efficiency of multiuser massive mimo-ofdm thz wireless systems with hybrid beamforming under inter-carrier interference,'' in \emph{2022 12th International Conference on Electrical and Computer Engineering (ICECE)}.\hskip 1em plus 0.5em minus 0.4em\relax IEEE, 2022, pp. 228--231.

\bibitem{THz_Beam_selection}
X.~Ma, Z.~Chen, Z.~Li, W.~Chen, and K.~Liu, ``Low complexity beam selection scheme for terahertz systems: A machine learning approach,'' in \emph{2019 IEEE International Conference on Communications Workshops (ICC Workshops)}, 2019, pp. 1--6.

\bibitem{yu2017hybrid}
X.~Yu, J.~Zhang, and K.~B. Letaief, ``Hybrid precoding in millimeter wave systems: How many phase shifters are needed?'' in \emph{GLOBECOM 2017-2017 IEEE Global Communications Conference}.\hskip 1em plus 0.5em minus 0.4em\relax IEEE, 2017, pp. 1--6.

\bibitem{nguyen2016hybrid}
D.~H. Nguyen, L.~B. Le, and T.~Le-Ngoc, ``Hybrid mmse precoding for mmwave multiuser mimo systems,'' in \emph{2016 IEEE international conference on communications (ICC)}.\hskip 1em plus 0.5em minus 0.4em\relax IEEE, 2016, pp. 1--6.

\bibitem{MM_wave_Beam_selection}
I.~Ahmed, M.~K. Shahid, and T.~Faisal, ``Deep reinforcement learning based beam selection for hybrid beamforming and user grouping in massive mimo-noma system,'' \emph{IEEE Access}, vol.~10, pp. 89\,519--89\,533, 2022.

\bibitem{self_HBF}
R.~U. Murshed, Z.~B. Ashraf, A.~H. Hridhon, K.~Munasinghe, A.~Jamalipour, and M.~F. Hossain, ``A cnn-lstm-based fusion separation deep neural network for 6g ultra-massive mimo hybrid beamforming,'' \emph{IEEE Access}, vol.~11, pp. 38\,614--38\,630, 2023.

\bibitem{ullah2023beyond}
M.~S. Ullah, R.~U. Murshed, and M.~F. Uddin, ``Beyond traditional beamforming: Singular vector projection techniques for mu-mimo interference management,'' \emph{arXiv preprint arXiv:2311.03741}, 2023.

\bibitem{ju2023142}
S.~Ju and T.~S. Rappaport, ``142 ghz multipath propagation measurements and path loss channel modeling in factory buildings,'' \emph{arXiv preprint arXiv:2302.12142}, 2023.

\bibitem{alkhateeb2015limited}
A.~Alkhateeb, G.~Leus, and R.~W. Heath, ``Limited feedback hybrid precoding for multi-user millimeter wave systems,'' \emph{IEEE transactions on wireless communications}, vol.~14, no.~11, pp. 6481--6494, 2015.

\bibitem{yuan2018hybrid}
H.~Yuan, N.~Yang, K.~Yang, C.~Han, and J.~An, ``Hybrid beamforming for mimo-ofdm terahertz wireless systems over frequency selective channels,'' in \emph{2018 IEEE Global Communications Conference (GLOBECOM)}.\hskip 1em plus 0.5em minus 0.4em\relax IEEE, 2018, pp. 1--6.

\end{thebibliography}
\vspace{12pt}
\end{document}